\documentclass[fleqn,12pt,twoside]{article}
\usepackage{espcrc1}

\usepackage{graphicx}
\usepackage[figuresright]{rotating}

\newcommand{\AmS}{{\protect\the\textfont2
  A\kern-.1667em\lower.5ex\hbox{M}\kern-.125emS}}

\title{Stable $uudd {\overline s}$ pentaquarks in the 
constituent quark model}

\author{Fl. Stancu\address[MCSD]{Universit\'{e} de Li\`ege, 
      Institut de Physique B.5,\\
        Sart Tilman, B-4000 Li\`ege 1, Belgium} 
        \thanks{e-mail : fstancu@ulg.ac.be}
	and
        D. O. Riska\address{Helsinki Institute of Physics\\
	  P.O.B. 64, 00014 University of Helsinki, Finland}
	\thanks{e-mail : riska@pcu.helsinki.fi} }

\begin{document}

\maketitle

\begin{abstract}
The stability of strange pentaquarks $uudd\overline{s}$
is studied in a constituent quark model based on a 
flavor-spin hyperfine interaction
between quarks. With this interaction model, which 
schematically represents the
Goldstone boson exchange interaction between constituent
quarks, the lowest lying strange
pentaquark is a $p-$shell state with positive parity.
The flavor-spin interaction lowers the energy of
the lowest $p-$shell state below that
of the lowest $s-$shell state, which has negative parity because
of the negative parity of the strange antiquark.
It is found that the strange pentaquark
is stable against strong decay provided that the strange
antiquark interacts by a fairly strong
spin-spin interaction with $u$ and $d$ quarks. This interaction has a form
that corresponds to $\eta$ meson exchange.
Its strength 
may be inferred from
the $\pi^0$ decay width of $D_s^*$ mesons.
\end{abstract}

\vspace{2cm}

Renewed interest in the existence of pentaquarks \cite{GI87,LI87}
has been raised by the 
recent observation of an $S = 1$ 
baryon resonance in photo-production of
kaon pairs on neutrons: 
$\gamma n \leftarrow K^+ K^- n$ \cite{NAKANO}.
This resonance has a peak at 1.54 $\pm$ 0.01 GeV/c$^2$ and a width,
which is less than 25 MeV/c$^2$. It may be interpreted as a strange
meson-baryon resonance or as a pentaquark of the form $uudd  \overline{s}$.

The expectation has been that stable pentaquarks should be likely
to exist in the heavy flavor sectors \cite{GI87,LI87,Sco,FS}, but
experimental searches have remained inconclusive \cite{Ashery,Ashery2}.
A constituent quark model study of pentaquark states of the form
$qqqq\bar s$, indicates that such states are unstable against strong
decay if the only interaction between the strange antiquark and
the light flavor quarks is the confining interaction \cite{Helminen}.
The prediction of a stable strange pentaquark with positive
parity at an the energy
close to that of the empirically found resonance was first made with
a chiral soliton model, in which it was classified as the lowest
state of the baryon antidecuplet multiplet \cite{Diak}. 
Here it is shown that once an attractive spin dependent hyperfine interaction
between the light flavor quarks and the strange antiquark
is introduced, stable
positive parity strange pentaquarks may also be accomodated
within the constituent quark model.

The originally proposed pentaquarks, which were introduced in the context of 
the conventional one-gluon exchange model for the
hyperfine interaction between constituent quarks, had negative parity, as
they represented states with all the light flavor quarks
as well as the strange
antiquark in their lowest $s-$states. Once the chromomagnetic
interaction is replaced by a spin and flavor dependent interaction,
with the form, which corresponds to a Goldstone boson  
exchange (GBE) interaction between quarks, 
\cite{GR96,GEB} the lowest lying pentaquarks will, however, have positive 
rather than negative parity \cite{FS}.   

The parity of the
pentaquark is given by P$\ =\ {\left({-}\right)}^{\mbox{L\ +\ 1}}$.  
Here, we take L = 1 and analyze the case 
where the subsystem of 
two $u$ and two $d$ quarks is in a state of
orbital symmetry ${\left[{31}\right]}_{O}$, which
thus carries the angular momentum L = 1. Although the kinetic 
energy of such
a state is higher than that of the orbitally symmetric state 
${\left[{4}\right]}_{O}$, an estimate based on a schematic
interaction model
 \cite{FS} shows that the
${\left[{31}\right]}_{O}$ symmetry should be the most favourable from
the point of view of stability against strong decays. 
In Ref. \cite{FS}
the antiquark was assumed to have heavy $c$ or $b$ flavor, 
and accordingly 
the interaction 
between a light 
quark and the heavy antiquark was neglected, which is justified
in the heavy quark limit. As the constituent mass of the
strange quark is not much larger than that of the
light flavor quarks, that approximation cannot be invoked
for strange pentaquarks. Below it is in fact shown that
the stable low lying strange pentaquarks only appear
if an interaction between ${\overline s}$ and the light quarks 
is included explicitly in the constituent quark model.

We shall employ the following schematic flavor-spin
interaction  between light quarks \cite{GR96}:  
\begin{equation}\label{CHIRAL}
{V}_{\chi }\ =\ -\ {C}_{\chi }\ \sum\limits_{ i\ <\ j}^{4} {\lambda }_{
i}^{ F} \cdot {\lambda }_{ j}^{ F} \ {\vec{\sigma }}_{i} \cdot {\vec{\sigma
}}_{j}.
\label{VCHI}
\end{equation}
Here $\lambda_i^F$ are Gell-Mann matrices for flavor $SU(3)$, 
and $\vec{\sigma}_i$ are the Pauli spin matrices. The constant
$C_{\chi}$ may be determined from the $\Delta$-N splitting
to be $C_{\chi}\cong$ 30 MeV
\cite{GR96}.
The interaction (\ref{VCHI}) is the simplest model for the
hyperfine interaction between quarks, which can describe
the empirical baryon spectrum in the constituent quark model
\cite{GR96}. It may be
interpreted as arising from pion and (mainly) two-pion
exchange, or more generally from exchange of the octet
of light pseudoscalar mesons (``Goldstone bosons'') and
vector mesons between the constituent quarks \cite{GEB,Graz}. 

The pion decay $D_s^*\rightarrow D_s\pi^0$ implies, by
$\pi^0 -\eta$ mixing, that $\eta$ mesons couple to strange
quarks and antiquarks \cite{LR}. It is then natural to
assume that there is an $\eta$ meson exchange interaction
between $\bar s$ antiquarks and light flavor quarks.
In particular it should lead to a spin-dependent interaction between 
the strange antiquark and the 4 light flavor quarks,
which is similar to (\ref{VCHI}).
This may be schematically be represented by the interaction: 
\begin{equation}\label{SPINSPIN}
{V}_{\eta }\ =\ \ {V}_{0} \ \sum\limits_{i}^{4}
{\vec{\sigma }}_{i} \cdot {\vec{\sigma}}_{\overline s}.
\end{equation}
Here $V_0$ is a constant, which should correspond to
the ground state matrix element of the spin-spin part
of the $\eta$ exchange interaction, but which here will
be taken to be a phenomenological constant.
The total hyperfine interaction is then
\begin{equation}\label{HYP}
V =\ \ V_{\chi} + V_{\eta}.
\end{equation}

The quark model values for the pseudovector coupling constant
between light flavor and $\eta$ mesons and strange
constituent quarks and $\eta$ mesons are
\begin{equation}
f_{\eta qq}={m_\eta\over 2\sqrt{3} f_\eta }g_A^q,\quad
f_{\eta ss}=-{m_\eta\over \sqrt{3} f_\eta }g_A^q.
\end{equation}
These expressions suggest that $f_{\eta qq}$ falls within
the range 1.25 -- 1.4 and
that
$f_{\eta ss}$ falls in the range -2.5 and -2.8.
Here $f_\eta$ = 112 MeV is the $\eta$ meson decay constant and $g_A^q$ is 
the axial coupling of the quark. The value of the latter
is expected to fall within the range 0.75 -- 1.0 
\cite{Weinberg}. 

The strength of the coupling between $\eta$ mesons and strange
constituent quarks may be derived from the known empirical
$\pi^0$ decay width of $D_s$ mesons, which is mediated by
$\eta$ mesons. This suggests that $f_{\eta ss} \sim -1.66$
\cite{LR}.  

For pseudoscalar mesons the coupling
to antiquarks has the same sign as that of quarks. Because of the
negative sign of the coupling of strange quarks to $\eta$ mesons
and the positive sign of the coupling of strange quarks to
light flavor constituent quarks, the potential coefficient
$V_0$ is expected to be positive.

An estimate of the $\eta$ meson exchange contribution to the strength 
of $V_\eta$ may be obtained  
from the expectation value of the 
radial part of the $\eta$ meson exchange interaction,
\begin{equation}
V_0 (r)= {f_{\eta qq} f_{\eta ss}\over 12\pi}
\{ {e^{-m_\eta r}\over r} - 4\pi {\delta(\vec r) \over m_\eta^2}\},
\end{equation}
in
the ground state of a quark-antiquark pair described by a 
harmonic oscillator
wave function
\begin{equation}
\varphi(\vec r) = ({\alpha^2\over \pi})^{3/4} e^{-\alpha^2 r^2/2},
\label{osc}
\end{equation}
where the parameter $\alpha$ may be adjusted to correspond to
a realistic wave function model.
This yields:
\begin{equation}
\langle V_0 \rangle  = m_\eta{f_{\eta qq} f_{\eta ss}\over 3\pi\sqrt{\pi}}
({\alpha\over m_\eta})^3\{{m_\eta^2\over 2\alpha^2}
-\sqrt{\pi}{m_\eta^3\over 4 \alpha^3}e^{m_\eta^2/4\alpha^2} 
\mathrm{erfc}({m_\eta\over 2\alpha}) - 1\}.
\label{est}
\end{equation}
With the values of the $\eta-$quark couplings above, this
expression yields values for $\langle V_0 \rangle$, which are
of the same order of magnitude as that of $C_\chi$, when the
baryon wavefunctions are compact, with matter radii less than
$1/m_\eta$. This condition is fullfilled for example by the model
in \cite{Gloz98}, for which the ground state wavefunction
may be approximated by a product of two oscillator functions
of the form (\ref{osc}) of the two Jacobi coordinates, with
$\alpha \simeq$ 650 MeV \cite{che}. With that value, and
with $f_{\eta qq}=$ 1.3 and $f_{\eta ss}=$ - 1.66, eqn.(\ref{est}) 
yields $\langle V_0 \rangle \sim$ 90 MeV. This number would be somewhat
reduced by the contribution from singlet pseudoscalar
exchange mechanisms like  $\eta'$-meson exchange \cite{Gloz98}.

For the construction of the wave functions for the pentaquark 
it is convenient to first consider the light quark $q^4$ subsystem. 
For this the Pauli
principle allows for the following two totally antisymmetric states with
${\left[{31}\right]}_{O}$ symmetry, written in the flavour-spin (FS) coupling
scheme \cite{FS,book}:
\begin{equation}\label{STATE1}
\left.{\left|{1}\right.}\right\rangle\ =\
\left({{\left[{31}\right]}_{O}{\left[{211}\right]}_{C}
{\left[{{1}^{4}}\right]}_{
OC}\
;\
{\left[{22}\right]}_{F}{\left[{22}\right]}_{S}{\left[{4}\right]}_{FS}}\right),
\end{equation}
\begin{equation}\label{STATE2}
\left.{\left|{2}\right.}\right\rangle\ =\
\left({{\left[{31}\right]}_{O}{\left[{211}\right]}_{C}
{\left[{{1}^{4}}\right]}_{
OC}\
;\
{\left[{31}\right]}_{F}{\left[{31}\right]}_{S}{\left[{4}\right]}_{FS}}\right).
\end{equation}
Asymptotically, a ground state baryon and a meson, into which a pentaquark can
split, would give ${\left[{3}\right]}_{O}\ \times  \ {\left[{2}\right]}_{O}\
=\ {\left[{5}\right]}_{O}\ +\ {\left[{41}\right]}_{O}\ +\
{\left[{32}\right]}_{O}$. By removing the antiquark, one can make the reduction
${\left[{41}\right]}_{O}\ \rightarrow  \ {\left[{31}\right]}_{O}\ \times
\ {\left[{1}\right]}_{O}$ or ${\left[{32}\right]}_{O}\ \rightarrow  \
{\left[{31}\right]}_{O}\ \times
\ {\left[{1}\right]}_{O}$. Thus, the symmetry ${\left[{31}\right]}_{O}$ of the
light quark subsystem is compatible with an L = 1 asymptotically separated
baryon plus meson system.\par

Each one of these two states, (\ref{STATE1}) or (\ref{STATE2}), has to 
be coupled to the antiquark state. 
The total angular momentum  $\vec{J}\ =\ \vec{L}\ +\ \vec{S}\ +\ 
\vec{s}_{\overline q} $, 
where
$\vec{L}$ and $\vec{S}$ are   
the angular momentum and spin of the
light flavor subsystem respectively and 
$\vec{s}_{\overline q} $ 
the spin of the antiquark 
$s$,
takes the values 
$J =\frac{1}{2}$ or $\frac{3}{2}$.
The resulting pentaquark states mix through  
the quark-antiquark spin-spin interaction (\ref{SPINSPIN}). 
Here we study the lowest case, $J =\frac{1}{2} $.  

For the stability problem the relevant quantity is
\begin{equation}\label{DELTA}
\Delta E =\ E(q^4\overline{q}) -\ E(q^3) -\
E(q\overline{q}),
\end{equation}
where $E(q^4\overline{q})$, $E(q^3)$ and $E(q\overline{q})$ are the 
masses of the pentaquark, of the ground state baryon and of the meson into
which the pentaquark decays, respectively.  
The multiquark system Hamiltonian used to calculate E is formed
of a kinetic energy term, 
a confining interaction  
and the hyperfine interaction (\ref{HYP}).

Consider first the contribution of (\ref{CHIRAL}) only.
In the $q^4$ subsystem the expectation value of (\ref{CHIRAL}) 
is $-28 \ C_{\chi}$ for
$\left|{\left.{1}\right\rangle}\right.$ and $-64/3 \ C_{\chi}$ for
$\left|{\left.{2}\right\rangle}\right.$. 
Thus $\left|{\left.{1}\right\rangle}\right.$ is the lowest state.
For the ground state $q^3$ system (the nucleon) the 
contribution is  $-14 \ C_{\chi}$. 
There is no such interaction in the $q {\overline q}$ pair.
We assume that the confinement energy roughly
cancels out in $\Delta E$. Then, the kinetic energy contribution 
to \,$\Delta E$ \, is \, $\Delta  KE\ =\ 5/4\ \hbar \omega$ in a harmonic 
oscillator model. It follows that for   
the state $\left|{\left.{1}\right\rangle}\right.$
the GBE contribution 
is $\Delta V_{\chi} =\ - 14 \ C_{\chi}$. With $\hbar\omega \approx$ 250 MeV,
determined from the N(1440)\, - \, N splitting \cite{GR96}, this would give
\begin{equation}\label{ESTIMATE}
\Delta  E\ =\ {\frac{5}{4}}\ \hbar \omega  \ -\ 14\ {C}_{\chi }\ =\ -\
107.5\, \mathrm{MeV}
\end{equation}
i.e. a substantial binding \cite{FS}. This is to be contrasted with the 
negative parity pentaquarks studied in Ref. \cite{GE98} 
within the same model, but where the lowest state has the orbital
symmetry $[4]_O$ so that one has
$\Delta  E\ =\ 3/4\ \hbar \omega  \ -\ 2\ {C}_{\chi }\ =\ 127.5$ MeV,
i.e. instability, in agreement with the detailed study made in \cite{GE98}.

The estimate (\ref{ESTIMATE})
is a consequence of the flavor dependence of the chiral interaction 
(\ref{CHIRAL}). 
For a specific spin state ${\left[{f}\right]}_{S}$, a schematic
color-spin interaction of type ${V}_{c\ m}\ =\ -\ {C}_{c\ m}\ 
\sum \ {\lambda
}_{i}^{c} \cdot{\lambda }_{j}^{c}\ {\vec{\sigma }}_{i} \cdot 
{\vec{\sigma }}_{j}$, which may represent the 
one gluon exchange interaction, does not
make a distinction between ${\left[{4}\right]}_{O}$ and
${\left[{31}\right]}_{O}$. Consequently, the ${\left[{31}\right]}_{O}$ 
state would appear to lie above the state ${\left[{4}\right]}_{O}$,
because of the kinetic energy term. The flavor-spin 
interaction (\ref{VCHI}) overcomes the excess of kinetic contribution in
${\left[{31}\right]}_{O}$ and generates a lower expectation value for
${\left[{31}\right]}_{O}$ than for ${\left[{4}\right]}_{O}$.

Consider now the total hyperfine interaction (\ref{HYP}). The matrix
elements of  $V_{\eta}$ of
(\ref{SPINSPIN}) are calculated with the functions
$\psi_1$ and $\psi_2$ given in the Appendix.
The interaction  (\ref{HYP}) now
leads to the following matrix to be diagonalized: 
\begin{equation}\label{MATRIX}
\renewcommand{\arraystretch}{2.5}\begin{array}{c|cc}
 &  \langle \psi_1| \psi_1 \rangle  & \langle \psi_2|\psi_2 \rangle \\
\hline
\langle \psi_1|\psi_1 \rangle  & -28 C_{\chi } & \frac{8}{\sqrt{2}} V_0 \\
\langle \psi_2|\psi_2 \rangle &  \frac{8}{\sqrt{2}} V_0 & 
-\frac{64}{3}C_{\chi } -4 V_0 
\\
\end{array}
\end{equation}
Note that the contribution of $V_{\eta}$ cancels out for the
state $\psi_1$ derived from $(\ref{STATE1})$.
Taking $C_{\chi}$ = 30 MeV, as mentioned above, the eigenvalues
of this matrix become
\begin{equation}\label{SOLUTION}
\langle V \rangle = -(740 + 2 V_0) \pm [10,000 - 400 V_0 + 36 V^2_0]^{1/2}.
\end{equation}
When $V_0$ = 0,
the lowest solution gives  $\langle V \rangle $ =
$\langle V_{\chi} \rangle $ = - 840 MeV, consistent with Ref. \cite{FS}.

\begin{figure}
\begin{center}
\includegraphics*{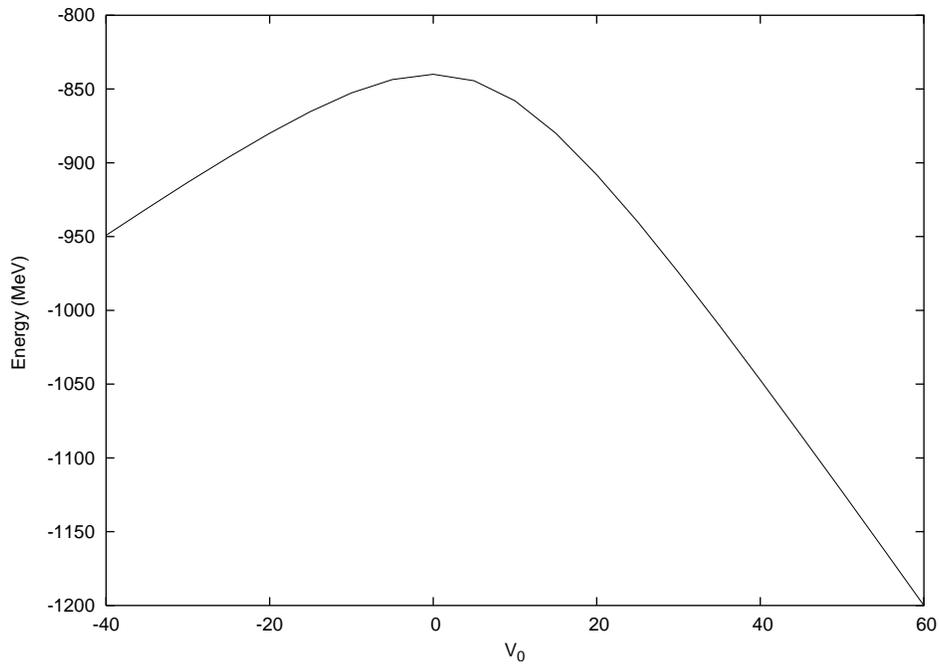}
\end{center}
\caption{\label{Fig1} The lowest solution of (\ref{SOLUTION}) as a function
of the parameter $V_0$. }
\end{figure}

In Fig. \ref{Fig1} the energy of the  lowest solution (\ref{SOLUTION})
is plotted as a function of the strength $V_0$. One can see that for a value of
$V_0 $ = $C_{\chi}$ the energy $E(q^4 {\overline q})$ can be lowered by 
about 130 MeV
with respect to the case $V_0$ = 0.
This implies a decrease by the same amount in $\Delta E$ of
(\ref{DELTA}) and hence a substantial increase in the stability 
of the system $uudd {\overline s}$. This may nevertheless not be sufficient
for ensuring stability. Actually estimates similar to those
of Ref. \cite{FS} containing only the GBE interaction \cite{GPP}
for which  
$E(q^3)$ = 969 MeV ,  give  $\Delta E$= 287 MeV. 
To get this value we have used $E(q {\overline q})$ = 793.6 MeV,
i. e. the average mass (M + M$^*$)/4 of the pseudoscalar K-meson mass M = 495
MeV 
and the vector K-meson mass M$^*$ = 893.1 MeV. 
To obtain a negative $\Delta E$ one 
needs $V_0 \approx $ 50 MeV, i. e. $V_0 \approx 5/3~ C_{\chi}$, as one can see 
from Fig. 1.

The estimate obtained from Eq.(6) above suggests that such strength of the 
spin-spin interaction between the light flavor quarks and the strange 
antiquark is quite plausible. While $\eta$ meson exchange is the most
obvious source of such an interaction, other mechanisms as two-kaon
exchange and $\eta'$ exchange should also contribute. 

The conclusion is that the stable strange pentaquarks with positive parity
can be accomodated by the constituent quark model, provided that: $1^0$ there 
is a flavor-spin dependent hyperfine interaction between the 4 light flavor
quarks, which is sufficiently strong for reversing the order of the lowest
states in the $s-$ and $p-$shells and that $2^0$ there is an at least as
strong spin-spin interaction between the light flavor and the strange
antiquark. The hyperfine chromomagnetic interaction between the quarks would 
in contrast not lead to stable pentaquarks with positive parity,
nor with negative.
While the the presence of a strongly flavor dependent hyperfine
interaction between constituent quarks originally was suggested by
phenomenological arguments alone \cite{GR96}, and in particular
by the requirement of reversal of normal ordering of the states in
the constituent quark model with 3 valence quarks, it has 
received further indirect support by recent QCD lattice calculations,
which show the same reversal of normal ordering for small
quark mass values \cite{KF}.

\vspace{1cm}

\centerline{\bf Acknowledgement}

DOR is indebted to L. Ya. Glozman for instructive correspondence
and Fl. S to M. Polyakov for a useful discussion.
Research supported by the Academy of Finland through grant 54038.

\vspace{1cm}

\centerline{\bf Appendix}
To calculate the matrix elements of the interaction (\ref{SPINSPIN})
first one has to couple the antiquark to the subsystem $q^4$. 
Then one has to decouple a $q {\overline s}$ pair from
the pentaquark system. One can work separately in the orbital, flavor, spin 
and color spaces. But as the interaction (\ref{SPINSPIN}) concerns
only the spin degree of freedom, the task is quite easy because
in the spin space the antiquark is on the same footing with the
quarks and the problem reduces to the usual
recoupling, via Racah coefficients. The only care must be taken of 
is the symmetry properties of the states.
Here we construct explicitly the flavor-spin part of the wave functions 
of the pentaquark.

Let us denote by $[f_{q^4}], [f_{q^3}], [f_{q^2}]$
and $[f]$ the partitions corresponding to the $q^4$, $q^3$,  
$q {\overline s}$ and  $q^4 {\overline s}$ respectively.
The corresponding spins are denoted by $J_q$, $j_1$, $S$ and $J$. 
For the two states (\ref{STATE1}) and (\ref{STATE2}) one has 
$[f_{q^4}]= [22]$, $J_q$ = 0 and 
$[f_{q^4}]= [31]$, $J_q$ = 1  respectively. The coupling to
the antiquark spin must therefore lead to the only common case
$[f] = [32]$, $J = 1/2$. The $q {\overline s}$ pair can have of course
$[f_{q^2}] = [2]$, $S$ = 1 or $[f_{q^2}] = [11]$, $S$ = 0.
Then the spin part of the wave function of the pentaquark reads
 
\begin{equation}\label{RECOUPLE}
[ \chi^{[f_{q^4}]}_{J_q} \chi^{[1]}_{1/2} ]^{~[f]}_{~JM} = 
\sum\limits_{S}{} [(2S + 1)(2 J_q + 1)]^{1/2}
W(j_1 \frac{1}{2} J \frac{1}{2}; J_q S) 
~[ \chi^{[f_{q^3}]}_{j_1} \chi^{[f_{q^2}]}_S  ]^{~[f]}_{~JM}.
\end{equation}
\noindent
In the recoupling one has however to keep track of the flavor-spin
symmetry of the subsystem of 4 identical quark.  This part of the 
wave function is 
symmetric, both in (\ref{STATE1}) and (\ref{STATE2}).
The flavor part of the wave function of $q^4$ should be specified but the 
recoupling with 
the antiquark does not have to be explicit, inasmuch as the interaction
(\ref{SPINSPIN}) is flavor independent. However the coupling to the
antiquark must give  the same quantum numbers 
($\lambda \mu$) = (11) in the flavor space, in both cases, otherwise
the scalar product cancels.

The two independent pentaquark flavor states 
associated with (\ref{STATE1}) are 

\begin{eqnarray}
{\phi }_{1} = (\renewcommand{\arraystretch}{0.5}
\begin{array}{c} $\fbox{1}\fbox{2}$ \\
$\fbox{3}\fbox{4}$\hspace{0mm} \end{array}
\times
{\phi}_{\overline s})^{(11)},
\end{eqnarray}

\begin{eqnarray}
{\phi }_{2} = (\renewcommand{\arraystretch}{0.5}
\begin{array}{c} $\fbox{1}\fbox{3}$ \\
$\fbox{2}\fbox{4}$\hspace{0mm} \end{array}
\times
{\phi}_{\overline s})^{(11)},
\end{eqnarray}
\noindent
where $\phi_{\overline s}$ is the flavor antiquark state.
Replacing the corresponding Racah coefficients in the 
relation (\ref{RECOUPLE}) 
the flavor-spin wave function of the pentaquark becomes
\begin{eqnarray}
|\psi_{1} \rangle = {|[22][1];[32] \rangle}_{1/2 M} \rangle & = & 
\frac{1}{\sqrt {2}} \{\phi_1  
~[ -\frac{1}{2} 
 [ \chi^{[21]}_{1/2} \chi^{[11]}_0 ]^{~[32]}_{~1/2M} 
+ \frac{\sqrt {3}}{2}
[ \chi^{[21]}_{1/2} \chi^{[2]}_1  ]^{~[32]}_{~1/2M}] ~ ] \\
\nonumber
&  & + \phi_2
~ [-\frac{1}{2} 
 [ \chi^{[21]}_{1/2} \chi^{[2]}_1 ]^{~[32]}_{~1/2M}] ~ ] \},
+ \frac{\sqrt {3}}{2}
 [ \chi^{[21]}_{1/2} \chi^{[2]}_1  ]^{~[32]}_{~1/2M}] ~ ] \},
\end{eqnarray} 
where in each row $\chi^{[21]}_{1/2}$ is associated with a
different Young tableau. 

The flavor-spin pentaquark  state constructed from (\ref{STATE2}) contains 
the following three independent flavor states

\begin{eqnarray}
{\phi }_{3} = (\renewcommand{\arraystretch}{0.5}
\begin{array}{c} $\fbox{1}\fbox{2}\fbox{3}$ \\
$\fbox{4}$\hspace{9mm} \end{array} 
\times
{\phi}_{\overline s})^{(11)},
\end{eqnarray}

\begin{eqnarray}
{\phi }_{4} =( \renewcommand{\arraystretch}{0.5}
\begin{array}{c} $\fbox{1}\fbox{2}\fbox{4}$ \\
$\fbox{3}$\hspace{9mm} \end{array}
\times
{\phi}_{\overline s})^{(11)},
\end{eqnarray}

\begin{eqnarray}
{\phi }_{5} = (\renewcommand{\arraystretch}{0.5}
\begin{array}{c} $\fbox{1}\fbox{3}\fbox{4}$ \\
$\fbox{2}$\hspace{9mm} \end{array}
\times
{\phi}_{\overline s})^{(11)}.
\end{eqnarray}
Then using these states and the recoupling (\ref{RECOUPLE}) with
corresponding Racah 
coefficients we obtain the pentaquark flavor-spin state 

\begin{eqnarray}
|\psi_{2} \rangle = {|[31][1];[32]\rangle}_{1/2 M} & = & 
\frac{1}{\sqrt {3}} \{ \phi_3  
~ [ \chi^{[3]}_{3/2} \chi^{[2]}_1  ]^{~[32]}_{~1/2M} \\
\nonumber
&  &+ \phi_4  
[~\frac{1}{2} 
 [ \chi^{[21]}_{1/2} \chi^{[2]}_1  ]^{~[32]}_{~1/2M}
+ \frac{\sqrt {3}}{2}
 [ \chi^{[21]}_{1/2} \chi^{[11]}_0  ]^{~[32]}_{~1/2M}] ~ \\
\nonumber
&  &+ \phi_5 
[\frac{1}{2} 
 [ \chi^{[21]}_{1/2} \chi^{[2]}_1  ]^{~[32]}_{~1/2M}
+ \frac{\sqrt {3}}{2}
[ \chi^{[21]}_{1/2} \chi^{[11]}_0  ]^{~[32]}_{~1/2M}] ~ \}.
\end{eqnarray}
\noindent
Again, in each row the function $\chi^{[21]}_{1/2}$ has a distinct Young
tableau. The explicit form of $q^3$ and $q^4$ flavor or spin states
associated with every Young tableau above can be found for example in 
Ref. \cite{book}.  

The wave functions $|\psi_{1} \rangle$ and  $|\psi_{2} \rangle$
are used to calculate the matrix elements of
the interaction (\ref{SPINSPIN}).


\begin{thebibliography}{9}

\bibitem{GI87} C. Gignoux, B. Silvestre-Brac and J.-M. Richard, 
Phys.Lett. {\bf B193} (1987) 323.
\bibitem{LI87} H.J. Lipkin, Phys.Lett. {\bf B195} (1987) 484.
\bibitem{NAKANO} T. Nakano et al. hep-ex/0301020
\bibitem{Sco} D. O. Riska and N. Scoccola, Phys. Lett. {\bf B299} (1993) 338
\bibitem{FS} Fl. Stancu, Phys. Rev. {\bf D58} (1998) 111501.
\bibitem{Ashery} E. M. Aitala et al., Phys. Rev. Lett. {\bf 81 } (1998) 44
\bibitem{Ashery2} E. M. Aitala et al., Phys. Lett. {\bf B448} (1999) 303
\bibitem{Helminen} C. Helminen and D. O. Riska, Nucl. Phys. {\bf A699} (2002)
624
\bibitem{Diak} D. Diakonov, V. Petrov and M. Polyakov, Z. Phys.
{\bf A359} (997) 305
\bibitem{GR96} L.Ya. Glozman and D.O. Riska, Phys. Rep. {\bf 268} (1996) 263.
\bibitem{GEB} D. O. Riska and G. E. Brown, Nucl. Phys. {\bf A653}
(1999) 251
\bibitem{Graz} R. F. Wagenbrunn et al., Nucl. Phys. {\bf A663\&A664}
(2000) 703c
\bibitem{LR} T. A. L\"ahde and D. O. Riska, Nucl. Phys {\bf 710} (2002) 
\bibitem{Weinberg} S. Weinberg, Phys. Rev. Lett. {\bf 67} (1991) 3473
\bibitem{che} C. Helminen, Phys. Rev. {\bf C59} (1999) 2829
\bibitem{Gloz98} L. Ya. Glozman et al., Phys. Rev. {\bf D58}
(1998) 094030 
\bibitem{book} Fl. Stancu, Group Theory in Subnuclear Physics, Oxford
University Press, Oxford 1996, chapter 4.
\bibitem{GE98} M. Genovese, J.-M. Richard, Fl. Stancu and S. Pepin, 
Phys.Lett. {\bf B425} (1998) 171.
\bibitem{GPP} L.Ya. Glozman, Z. Papp and W. Plessas, Phys. Lett. {\bf B381}
 (1996) 311.
\bibitem{KF} S. Dong et al, hep-ph/0306199

\end{thebibliography}
\end{document}